\documentclass[conference]{IEEEtran}
\IEEEoverridecommandlockouts

\ifCLASSINFOpdf
\else
\fi
\usepackage{xfrac}
\usepackage{amsmath}
\usepackage{multirow}
\usepackage{color}
\usepackage{amssymb}
\usepackage{cite}
\DeclareMathOperator*{\argmin}{argmin}

\usepackage{subcaption}

\usepackage{stfloats}
\begin{document}
\bstctlcite{IEEEexample:BSTcontrol}

\title{Error Probability Analysis of Non-Orthogonal Multiple Access with Channel Estimation Errors}
%
\author{\IEEEauthorblockN{
Ferdi KARA, Hakan KAYA}
\IEEEauthorblockA{
Wireless Communication Technologies Laboratory (WCTLab) \\
Department of Electrical and Electronics Engineering\\
Zonguldak Bulent Ecevit University\\
Zonguldak, TURKEY 67100\\
Email: \{f.kara,hakan.kaya\}@beun.edu.tr}
\thanks{This work is supported by Zonguldak Bulent Ecevit University with grant no: 2019-75737790-01.
}}

%
%

\markboth{International Conference on Communications}%
{Kara \MakeLowercase{and} Kaya: Spatial Multiple Access (SMA): Enhancing performances of MIMO-NOMA systems}
%
\maketitle
\begin{abstract}
Non-orthogonal multiple access (NOMA) is very promising for future wireless systems thanks to its spectral efficiency. In NOMA schemes, the effect of imperfect successive interference canceler (SIC) has dominant effect on the error performances. In addition to this imperfect SIC effect, the error performance will get worse with the channel estimation errors just as in all wireless communications systems. However, all literature has been devoted to analyze error performance of NOMA systems with the perfect channel state information (CSI) at the receivers which is very strict/unreasonable assumption. In this paper, we analyze error performance of NOMA systems with imperfect SIC and channel estimation errors, much more practical scenario. We derive exact bit error probabilities (BEPs) in closed-forms. All theoretical analysis is validated via computer simulations. Then, we discuss optimum power allocation for user fairness in terms of error performance of users and propose a novel power allocation scheme which achieves maximum user fairness.
\end{abstract}
%
\IEEEpeerreviewmaketitle
\section{Introduction}
%
%
%
%
Non-Orthogonal Multiple Access (NOMA) technique is one of the most strong candidates for future wireless systems, especially for Massive Machine Type Communication (MMTC) since it provides high spectral efficiency and allows ultra dense network \cite{Shirvanimoghaddam2017}. In NOMA schemes, users share same resource block with different power allocation coefficients. Thus, the users can be served at the same time and massive connections can be possible in a cell. Due to sharing same resource blocks, the users suffer from inter-user-interference. Nevertheless, this interference can be eliminated by successive interference canceler (SIC) at the receivers.

The usage of NOMA in future wireless systems has been proposed in \cite{Saito2013}, and its superiority to Orthogonal multiple access (OMA) schemes has been proved in terms of achievable/sum rate. Then, NOMA has been analyzed for outage probability and its performance gain in also outage performance has been presented\cite{Ding2014}. Thanks to these performance advantages, NOMA has attracted tremendous attention from both academia and industry\cite{Papers2017}. Moreover, since NOMA can easily be implemented in physical layer, a great deal of studies investigate  NOMA integration with other physical layer techniques \cite{Vaezi2019a} such as cooperative communication\cite{Liu2015b,Kara2019a}, multiple-input-multiple-output (MIMO)\cite{Ding2016a,Senel2019}, index modulation\cite{Kara2018,Kara2019}, visible light communication\cite{Marshoud2015} and reconfigurable intelligent surfaces\cite{Hou2019}.

On the other hand, the main drawback of the NOMA is the error performance due to inter-user-interference. Thus, a remarkable number of studies investigate NOMA in terms of error performance and bit/symbol/block error probabilities are derived for NOMA in various schemes according to  fading channels, number of users and number of transmit/receive antenna \cite{Kara2018d,Yeom2019,Assaf2019,Bariah2018,Kara2019b,Kara2020,Zheng2019}. Although, these studies consider imperfect SIC to characterize a practical scenario, all of them assume that perfect channel state information (CSI) is available at the receivers. However, considering channel estimation techniques, this assumption is too strict to be reasonable, hence it should be relaxed. To this end, in this paper, we analyze error performance of NOMA schemes when the imperfect SIC and channel estimation errors both exist. We derive exact bit error probabilities (BEPs) in closed-forms. To the best of the authors' knowledge, this is the first study which investigates error performance of NOMA schemes with imperfect SIC and CSI errors.

The rest of paper is organized as follows. In section II, the system and channel models are introduced and the detections at the users are given. In section III, the BEP expressions are derived for users. Then, the validation of the analytical analysis  via computer simulations is presented in Section IV. Moreover, the optimum power allocation for user fairness is discussed in Section V. Finally, section V discusses the results and concludes the paper.

\section{System Model}
We consider a two-user\footnote{Although more than two users can be implemented in NOMA, it is limited by two since increasing the number of users causes more inter-user-interference and users will have worse error performance. Thus, it is considered to be two users in also 3GPP standards\cite{PP2016}.} downlink NOMA scheme where a base station (BS) and two mobile users (i.e., $UE_i,\ i=1,2$) are located. One of the users is close to (in terms of Euclidean distance) BS, thus it is called near user ($UE_1$) and the other is far user ($UE_2$). The channel coefficient between each node follows\footnote{In the following of this paper the notation used are as follow. $CN(\mu,\sigma)$ is a complex Gaussian distribution which has independent real and imaginary random variables with the $\mu$ mean and the $\frac{\sigma}{2}$ variance. We use $\left|.\right|$ for the absolute value of a scalar/vector and $P_r(A)$ denotes the probability of the event $A$ whereas $P_r(A|B)$ is the probability of the event $A$ under the condition that $B$ has already occurred.} $CN(0,\sigma^2_{i})$, $i=1,2$ where $\sigma^2_1\geq\sigma^2_2$.  All nodes are equipped with single antenna. The BS implements a superposition coding for the symbols of the users and transmits it to the users simultaneously. Thus, the received signals by the users are given
\begin{equation}
y_i=\sqrt{P_s}\left(\sqrt{\alpha}x_1+\sqrt{\left(1-\alpha\right)x_2}\right)h_i+n_i, \ i=1,2,
\end{equation}
where $P_S$ is the transmit power of the BS. $x_i$ is the base-band modulated symbol\footnote{We assume that the symbols of both users are modulated by BPSK.} of the $UE_i$. $h_i$ is the flat fading channel coefficient from BS to $UE_i$. $n_i$ is the additive Gaussian noise (AGN) at the receiver of $UE_i$ and follows $CN(0,N_0)$. In (1), $\alpha$ denotes the power allocation coefficient. Since the $UE_1$ has better channel condition, it is assumed to be $\alpha<0.5$.

\subsection{Detection at the users}
\subsubsection{Far User ($UE_2$)}
The symbols of $UE_2$ have more power (i.e., $\sqrt{1-\alpha}$), thus $UE_2$ implements maximum likelihood (ML) detection by pretending $x_1$ symbols as noise. The ML detection at the $UE_2$ is given as
\begin{equation}
\hat{x}_2=\argmin_{n}{\left|y_2-\sqrt{\left(1-\alpha\right) P_s}x_{2,n}\hat{h}_2\right|^2},
\end{equation}
where $x_{2,n}$ denotes the $n$ th point within the constellation of $x_2$ symbols. $\hat{h}_2$ is the estimated channel coefficient at the $UE_2$ and it is given by $\hat{h}_2=h_2-\epsilon$ where $\epsilon$ follows $CN(0,\delta^2)$.
\subsubsection{Near User ($UE_1$)}
Since the $x_1$ symbols have less power, it is not possible to detect them directly as $x_2$ symbols. Thus, $UE_1$ should implement successive interference canceler where $x_2$ symbols are detected firstly like (2), then the detected $\hat{x}_2$ symbols are subtracted from received signal and finally the $x_1$ symbols are detected by ML. The detection procedure at $UE_1$ is given as
\begin{equation}
\hat{x}_1=\argmin_{n}{\left|y_1^+-\sqrt{\alpha P_s}x_{1,n}\hat{h}_1\right|^2},
\end{equation}
where
\begin{equation}
y_1^+=y_1-\sqrt{\left(1-\alpha\right) P_s}\hat{x}_{2}\hat{h}_1,
\end{equation}
and
\begin{equation}
\hat{x}_2=\argmin_{n}{\left|y_1-\sqrt{\left(1-\alpha\right) P_s}x_{2,n}\hat{h}_1\right|^2}.
\end{equation}
In (3)-(5), $x_{1,n}$ is the $n$ th point within the constellation of $x_1$ symbols. $\hat{h}_1$ is the estimated channel coefficient at the $UE_1$ and it is defined as $\hat{h}_1=h_1-\epsilon$.
\section{Bit Error Probability (BEP) Analysis}
Since only ML detection is implemented at $UE_2$, the erroneous detection at the $UE_2$ depends on whether the AGN is greater than the energy of the symbol or not. However, BS implements a superposition coding for $x_1$ and $x_2$ symbols and transmits this total symbol to the users. Thus, the energies of superposition-coded symbols change according to which $x_1$ and $x_2$ symbols are superimposed. Considering BPSK is used for symbols, two different energy levels can be produced with equal probabilities. In addition, the channel estimation errors should also be considered for the erroneous detection probability. Based on these discussions, the conditional BEP of $UE_2$ is given by
\begin{equation}
\begin{split}
   P_2(e|_{h_2})=&\frac{1}{2}P\left(n_2+\sqrt{P_s}\left(\sqrt{1-\alpha}+\sqrt{\alpha}\right)\epsilon> \right.\\
   &\left.\sqrt{P_s}\left(\sqrt{1-\alpha}+\sqrt{\alpha}\right)h_2\right)\\ &\frac{1}{2}P\left(n_2+\sqrt{P_s}\left(\sqrt{1-\alpha}-\sqrt{\alpha}\right)\epsilon> \right. \\
   &\left.\sqrt{P_s}\left(\sqrt{1-\alpha}-\sqrt{\alpha}\right)h_2\right)
    \end{split}
\end{equation}
With some simplifications and algebraic manipulations, it is determined as
\begin{equation}
    P_2(e|_{\gamma_2})=\sum_{k=1}^2{\frac{1}{2}Q\left(\sqrt{2\gamma_{2,k}^{'}}\right)}
\end{equation}
where $\gamma_{2,k}^{'} =\frac{\beta_k\rho_s\gamma_2}{\beta_k\delta^2\rho_s+1}$, $\beta_k=[1+\sqrt{\alpha-\alpha^2},1-\sqrt{\alpha-\alpha^2}]$, $\rho_s=\sfrac{P_s}{N_0}$ and $\gamma_2\triangleq\left|h_2\right|^2$ are defined. The $\gamma_2$ follows exponential distributions. When the conditional BEP in (7) is averaged over instantaneous $\gamma_2$, with the aid of \cite{Alouini1999}, the average BEP (ABEP) of $UE_2$ is derived as
\begin{equation}
    P_2(e)=\sum_{k=1}^{2}{\frac{1}{4}\left(1-\sqrt{\frac{\beta_k\rho_s\sigma^2_2}{\beta_k\delta^2\rho_s+1}}\right)}.
\end{equation}

On the other hand, the BEP of $x_1$ symbols should be considered for two cases (i.e., correct SIC and erroneous SIC). Thus, with the law of probability, the ABEP of $UE_1$ is given by
\begin{equation}
    P_1(e)= P_{SIC}P_1(e|_{error})+\left(1-P_{SIC}\right)P_1(e|_{correct})
\end{equation}
where $P_{SIC}$ is the ABEP for $x_2$ symbols detection at the $UE_1$. $P_1(e|_{error})$ and $P_1(e|_{correct})$ denote the ABEP of $x_1$ symbols when the $x_2$ symbols are detected erroneously and correctly, respectively.

In order to obtain $P_{SIC}$, we repeat the steps between (6) and (8) for $h_1$, then it is derived as
\begin{equation}
    P_2(e)=\sum_{k=1}^{2}{\frac{1}{4}\left(1-\sqrt{\frac{\beta_k\rho_s\sigma^2_1}{\beta_k\delta^2\rho_s\sigma^2_1+1}}\right)}.
\end{equation}

If the SIC is implemented erroneously at the $UE_1$, this means that wrongly estimated $\hat{x}_2$ symbols will be subtracted from the received signal $y_1$ as given in (4). Without loss of the generality, we can consider this wrongly-subtraction as noise for detection of $x_1$ symbols. Therefore, the noise term will be greater than the energy of $x_1$ symbols thereby the ABEP in this case is close to the worst case as
\begin{equation}
P_1(e|_{error})\cong\frac{1}{2}
\end{equation}

As the second case, if correct SIC is implemented at the $UE_1$, only $x_1$ symbols will remain after subtraction. Nevertheless, due to the SIC operation, the effect of channel estimation error will be increased. Hence, the conditional BEP of $x_1$ symbols in case correct SIC is given as
\begin{equation}
    P_1(e|_{h_1\cap correct})=P\left(n_1+\sqrt{P_s}\epsilon>\sqrt{\alpha P_s}h_1\right)
  \end{equation}
 and it is obtained as
 \begin{equation}
    P_1(e|_{\gamma_1\cap correct})=Q\left(\sqrt{2\gamma_{1}^{'}}\right),
  \end{equation}
$\gamma_{1}^{'} =\frac{\alpha\rho_s\gamma_1}{\delta^2\rho_s+1}$ and $\gamma_1\triangleq\left|h_1\right|^2$ are defined. Just as (8), with the aid of \cite{Alouini1999}, the ABEP of $x_1$ symbols in correct SIC case is derived as
\begin{equation}
    P_1(e|_{correct})=\frac{1}{2}\left(1-\sqrt{\frac{\alpha\rho_s\sigma^2_1}{\delta^2\rho_s+1}}\right).
\end{equation}

Finally, substituting (10), (11) and (14) into (9), the ABEP for $x_1$ symbols derived as in (15) (see the top of the page).
\begin{figure*}[t]
\begin{equation}
  P_1(e)= \sum_{k=1}^{2}{\frac{1}{8}\left(1-\sqrt{\frac{\beta_k\rho_s\sigma^2_1}{\beta_k\delta^2\rho_s+1}}\right)}+\left\{1-\sum_{k=1}^{2}{\frac{1}{4}\left(1-\sqrt{\frac{\beta_k\rho_s\sigma^2_1}{\beta_k\delta^2\rho_s+1}}\right)}\right\} \frac{1}{2}\left(1-\sqrt{\frac{\alpha\rho_s\sigma^2_1}{\delta^2\rho_s+1}}\right)
\end{equation}
    \hrulefill
\end{figure*}
\section{Numerical Results}

In this section, we present Monte Carlo simulations to validate theoretical analysis. In all figures, simulations are presented by markers and theoretical curves are denoted by lines. All simulation results are obtained from $10^7$ channel realizations.

In Fig. 1 and Fig. 2, we present error performances of users with the change of transmit SNR for power allocation coefficients $\alpha=0.1$  and $\alpha=0.2$, respectively. In both figures, we assume that $\sigma^2_1=10dB$ and $\sigma^2_2=0dB.$ The results are presented for five different channel estimation error effects as $\delta=0$ (perfect CSI) and $\delta=0.01, 0.02, 0.05, 0.1$. It is noteworthy that derived expressions match perfectly with simulations. In addition, as expected, with the increase of channel estimation errors (i.e., $\delta$) both users' performance decrease. Especially, in high SNR region, an error floor may be observed. Besides, if we compare Fig. 1 and Fig. 2, we can easily see that increasing $\alpha$ provides performance gain for $UE_1$  whereas a performance decay occurs for $UE_2$. This is also an expected results since the higher $\alpha$ means that the higher and lower powers are allocated to symbols of $UE_1$ and $UE_2$, respectively. Nevertheless, the gain in the $UE_1$ 's performance cannot be always-observed with the increase of $\alpha$ due to the SIC operation.
 \begin{figure}[t]
\begin{subfigure}{.5\textwidth}
  \centering
  \includegraphics[width=8.5cm,height=5.7cm]{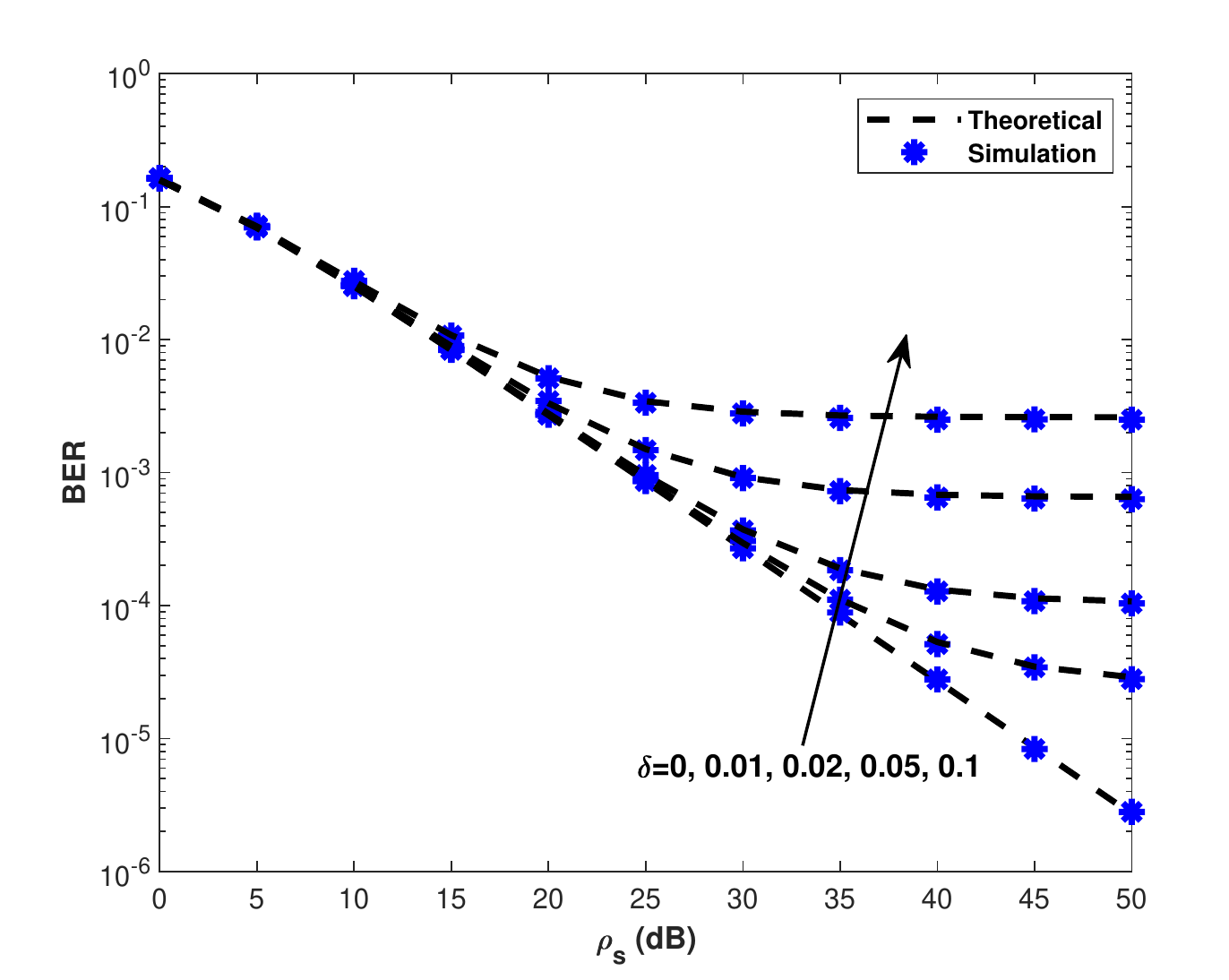}
  \caption{$UE_1$}
  \label{fig:sub-first}
\end{subfigure}
\begin{subfigure}{.5\textwidth}
  \centering
  \includegraphics[width=8.5cm,height=5.7cm]{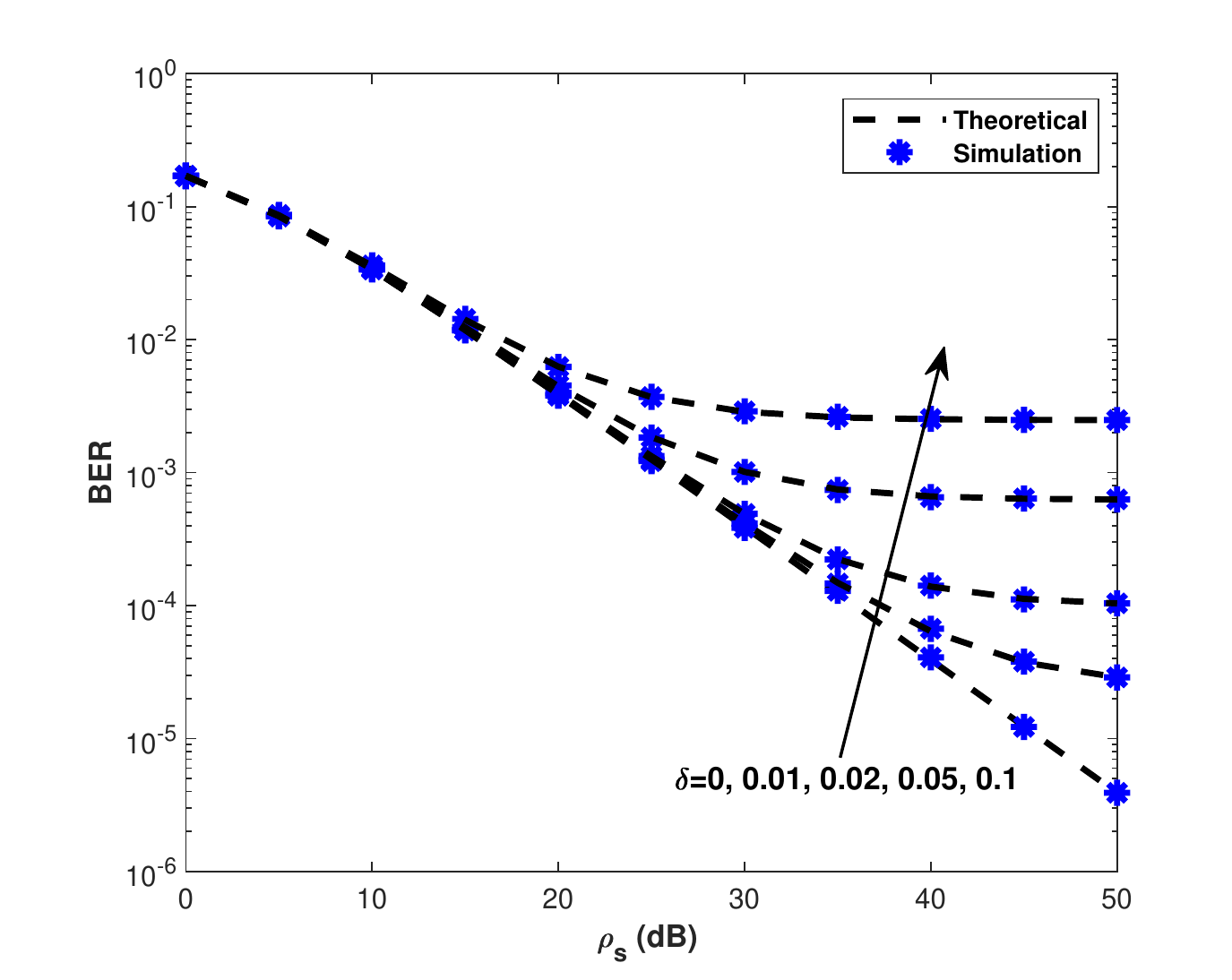}
  \caption{$UE_2$}
  \label{fig:sub-second}
\end{subfigure}
\caption{Error performance of NOMA vs. transmit SNR ($\rho_s$) with channel estimation errors ($\delta=0, 0.01, 0.02, 0.05, 0.1$) when $\sigma_{1}^2= 10dB$, $\sigma_{2}^2=0dB$, $\alpha=0.1$}
\label{fig:fig}
\end{figure}
\begin{figure}[t]
\begin{subfigure}{.5\textwidth}
  \centering
  \includegraphics[width=8.5cm,height=5.7cm]{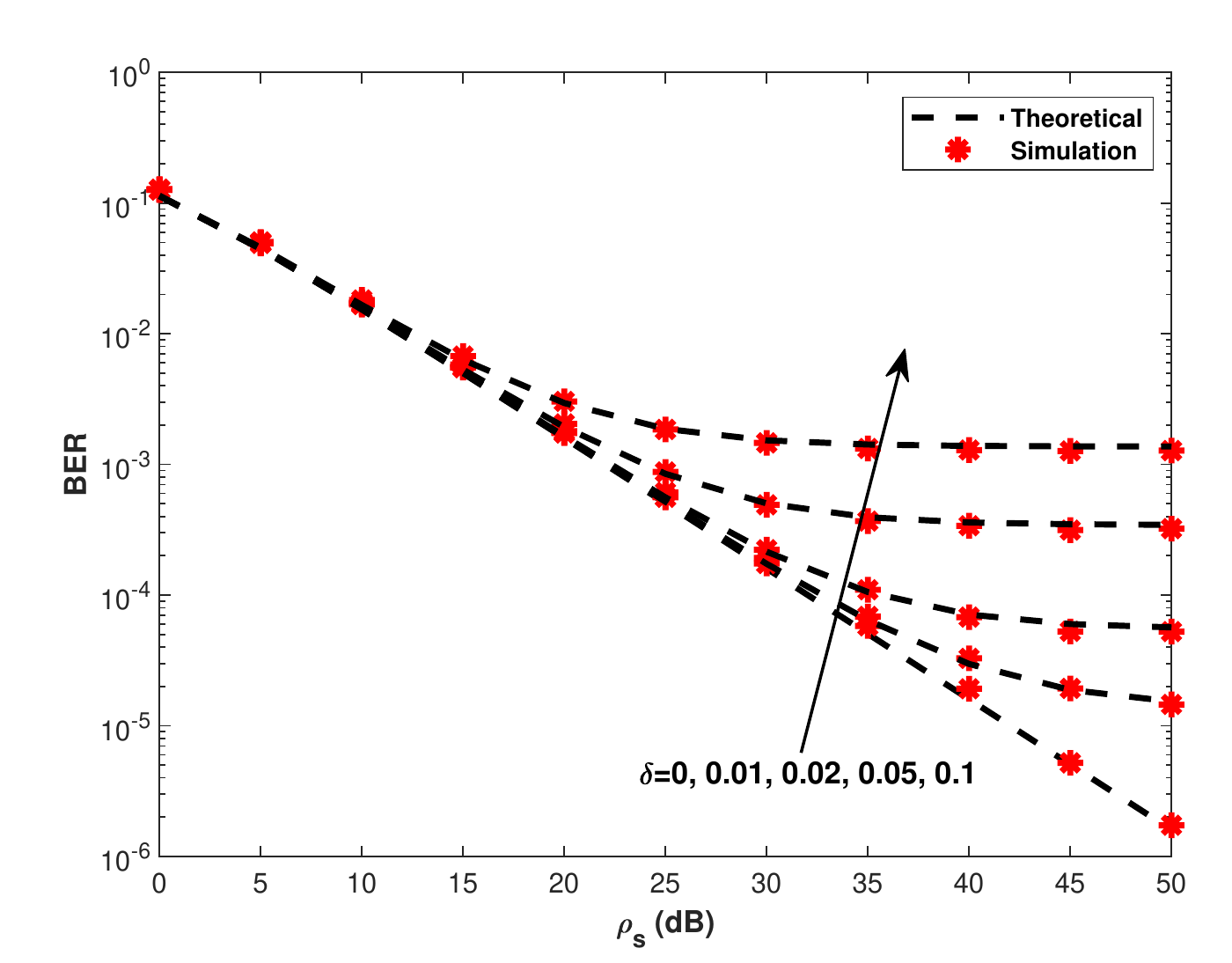}
  \caption{$UE_1$}
  \label{fig:sub-first}
\end{subfigure}
\begin{subfigure}{.5\textwidth}
  \centering
  \includegraphics[width=8.5cm,height=5.78cm]{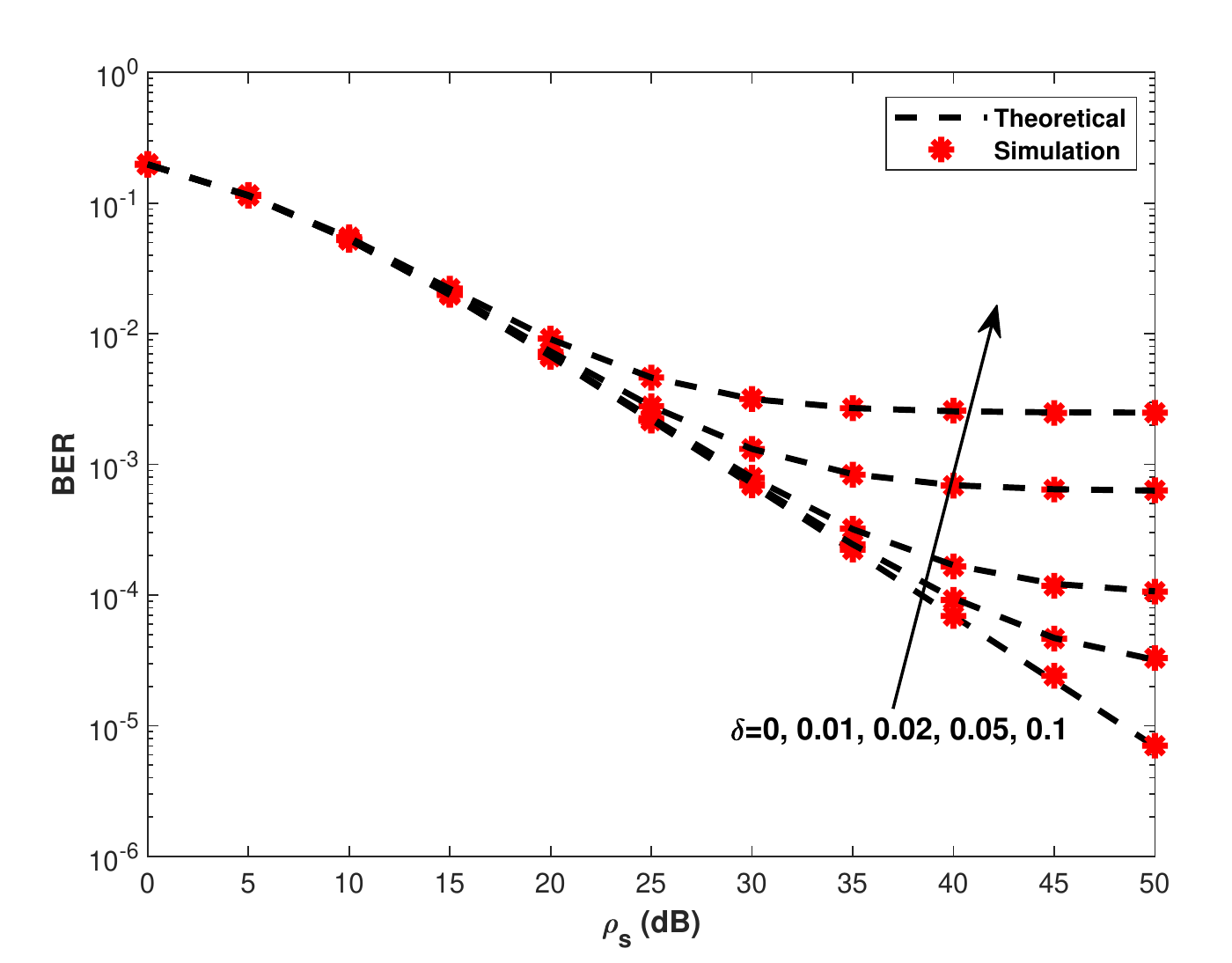}
  \caption{$UE_2$}
  \label{fig:sub-second}
\end{subfigure}
\caption{Error performance of NOMA vs. transmit SNR ($\rho_s$) with channel estimation errors ($\delta=0, 0.01, 0.02, 0.05, 0.1$) when $\sigma_{1}^2= 10dB$, $\sigma_{2}^2=0dB$, $\alpha=0.2$ }
\label{fig:fig}
\end{figure}
In order to reveal the effect of power allocation on the performances of users, we present error performances of users with the change of $\alpha$ in Fig. 3. The results are provided for transmit SNR $\rho_s=30dB$ and the channel estimation errors are assumed to be $\delta=0$ (perfect CSI) and $\delta=0.01, 0.02, 0.05, 0.1$. The channel conditions are assumed to be same with previous figures. One can easily see that increase of $\alpha$ always causes a decay in error performance of $UE_2$. On the other hand, it firstly provides a gain in error performance of $UE_1$, but too much increase in $\alpha$ causes a decay in performance of $UE_1$. This can be explained as follow. With the increase of $\alpha$, erroneous detection of $UE_2$ symbols at the $UE_1$ increases, thus these erroneous detected $UE_2$ symbols cause erroneous SIC operation (wrongly detected $x_2$ symbols are subtracted from the receives signal) and the error performance of the $UE_1$ becomes worse. Furthermore, the effect of imperfect CSI will change on the performance of users according to the power allocation. For instance, when relatively lower power allocation is implemented (e.g., $\alpha<0.25$), the error performance of $UE_2$ is affacted more with the increase of channel estimation errors (i.e., $\delta$). Nevertheless, $UE_2$ will have almost the same error performance when $\alpha\geq0.25$. The same discussion is valid for $UE_1$ since the error performance of $UE_1$ is dominated by SIC operation (same with the error performance of $UE_2$).
   \begin{figure}[!t]
   \centering
     \includegraphics[width=8.5cm,height=5.7cm]{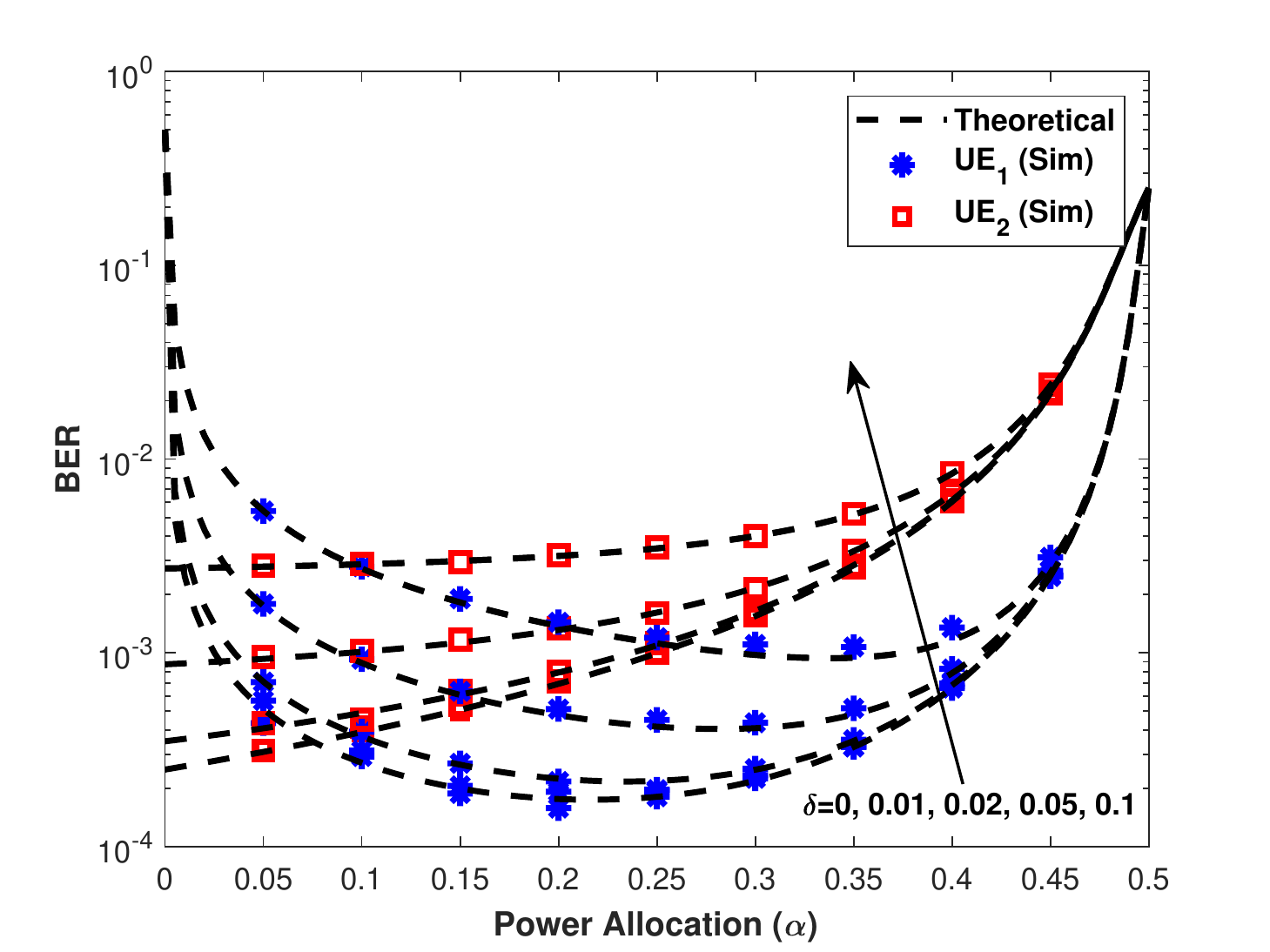}
    \caption{Error performance of NOMA vs. power allocation ($\alpha$) with channel estimation errors ($\delta=0, 0.01, 0.02, 0.05, 0.1$) when $\sigma_{1}^2= 10dB$, $\sigma_{2}^2=0dB$, $\rho_s=030dB$}
    \label{fig3}
 \end{figure}
 \section{Optimum Power Allocation}
 Based on extensive simulations in the previous section, we reveal that the power allocation has dominant effect on the error performances of users. Thus, the power allocation coefficient should be carefully chosen not to cause a user unfairness. To this end, we propose the optimum power allocation as
 \begin{equation}
 \begin{split}
    \alpha^*= &\min\{\max\{P_1(e),P_2(e)\}\}.\\
    &\text{s.t} \ \alpha^*<0.5
 \end{split}
 \end{equation}
With the proposed power allocation in (16), the ABEPs of users are calculated and the power allocation is chosen as the optimum value which minimizes the maximums of the ABEPs of the users. Therefore, none of the users performs much worse than the other so that the user fairness is achieved. In (16), $\alpha^*<0.5$ constraint should be satisfied, otherwise, the symbols of the users cannot be detected with the given SIC order. To evaluate the proposed power allocation, we present proportional fairness (PF) index\footnote{We hereby note that PF index of any performance metrics have the same meaning for $\kappa$ and $\sfrac{1}{\kappa}$. It only defines which user has better performance. For instance, $2$ and $0.5$ have the same meanings in terms of PF index} (i.e., $\sfrac{P_1(e)}{P_2(e)}$) for users in Fig 4. when users have same channel conditions (i.e., $\sigma^2_1=\sigma^2_2=0dB$). As seen from Fig. 4, the proposed power allocation outperforms fixed power allocation strategies. For instance, at $20dB$ when $\delta=0.05$, proportional fairness indexes are $8.56$, $3.68$ and $1.5$ for $\alpha=0.1$, $\alpha=0.2$ and proposed power allocation, respectively. This means that $UE_1$ has $8.56$ and $3.68$ times worse performance  than $UE_2$ for  $\alpha=0.1$ and $\alpha=0.2$. On the other hand, the proposed power allocation provides almost same performances for users and this proves the effectiveness of that for user fairness. Moreover, the proposed power allocation scheme is more robust to channel estimation errors in terms of PF. User fairness is increased even if the channel estimation errors increase whereas it gets worse for fixed power allocations.
  \begin{figure}[!t]
   \centering
     \includegraphics[width=8.5cm,height=5.7cm]{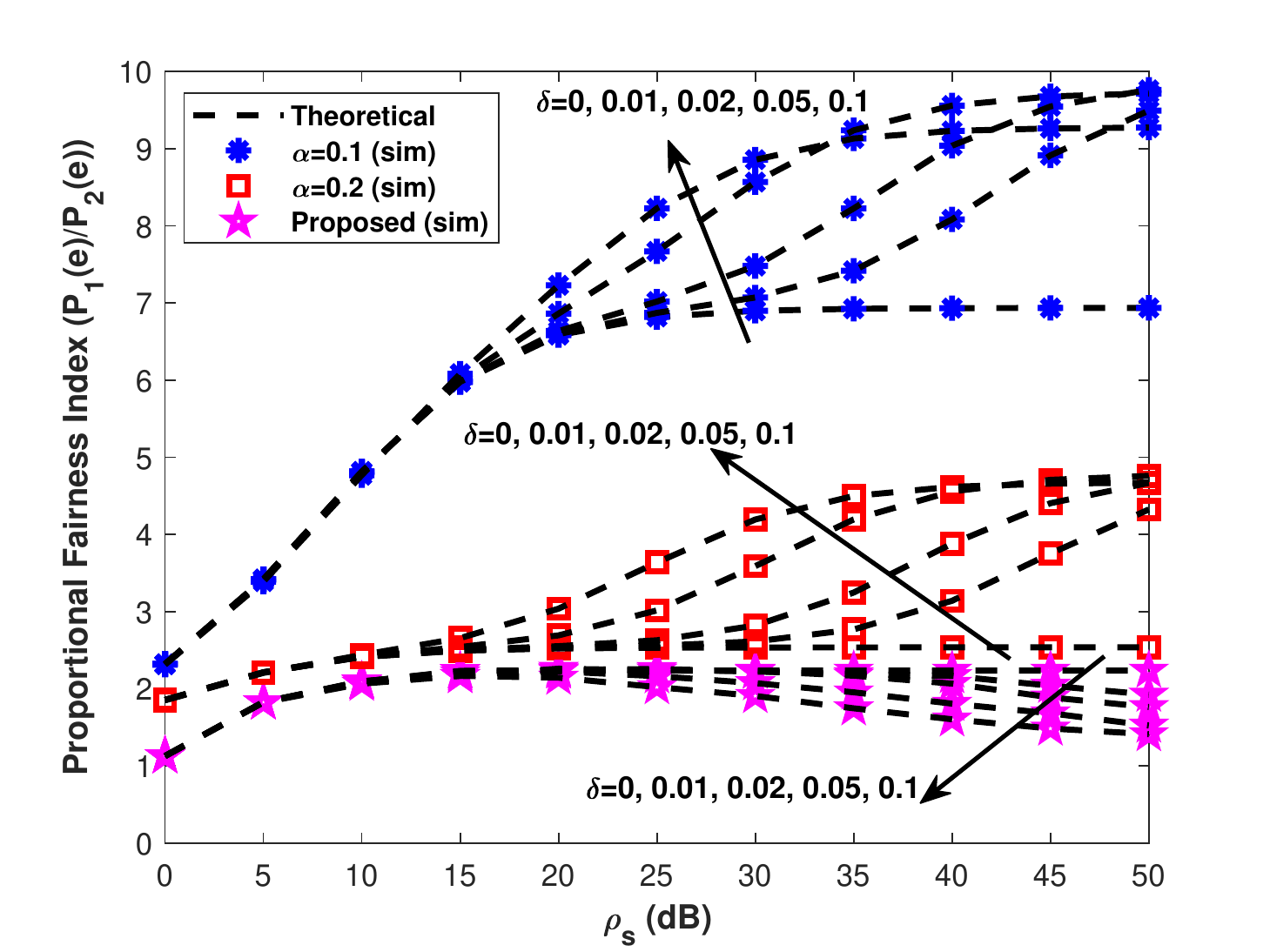}
    \caption{Proportional fairness index ( $\sfrac{P_1(e)}{P_2(e)}$) vs. transmit SNR ($\rho_s$) with channel estimation errors ($\delta=0, 0.01, 0.02, 0.05, 0.1$) when $\sigma_{1}^2=\sigma_{2}^2=0dB$}
    \label{fig4}
 \end{figure}
\section{Conclusion}
In this paper, we investigate the error performance of NOMA schemes in the presence of channel estimation errors in addition to imperfect SIC. We derive exact ABEP expressions in the closed-forms and all anlyses are validated via computer simulations. Moreover, we discuss optimum power allocation considering user fairness in terms of error performances. To the best of the authors' knowledge, error analysis of any NOMA schemes with channel estimation errors is firstly conducted in this paper. Thus, the analysis in this paper could be further extended for any NOMA schemes such as cooperative-NOMA, MIMO-NOMA, etc. These are seen as future works.

%
\ifCLASSOPTIONcaptionsoff
  \newpage
\fi
%
\IEEEtriggeratref{9}
\bibliographystyle{IEEEtran}
\bibliography{noma_ipSIC}
%
\end{document}